\documentclass[preprints,article,accept,pdftex,moreauthors]{Definitions/mdpi}
\firstpage{1} 
\pubvolume{1}
\issuenum{1}
\articlenumber{0}
\pubyear{2023}
\copyrightyear{2023}
\datereceived{} 
\dateaccepted{} 
\datepublished{} 
\hreflink{https://doi.org/} 
\pdfoutput=1

\Title{Two-Dimensional Strain Mapping with Scanning Precession Electron Diffraction: An Investigation into Data Analysis Routines}

\TitleCitation{Title}

\Author{Phillip Crout$^{1}$\orcidA{}, Dipanwita Chatterjee$^{2}$, Ingeborg Nævra Prestholdt$^{2}$, Tor Inge Thorsen$^{2}$, P. A. Midgley$^{1}$ and Antonius T. J. van Helvoort$^{2}$}

\AuthorNames{Firstname Lastname, Firstname Lastname and Firstname Lastname}

\AuthorCitation{Lastname, F.; Lastname, F.; Lastname, F.}

\address{%
$^{1}$ \quad Department of Materials Science and Metallurgy, University of Cambridge, 27 Charles Babbage Road, Cambridge, CB3 0FS, UK\\
$^{2}$ \quad Department of Physics, Høgskoleringen, Norwegian University of Science and Technology, Trondheim, Norway}

\corres{Correspondence: a.helvoort@ntnu.no}

\abstract{Scanning precession electron diffraction (SPED) is a powerful technique for investigating strain. While extensive literature exists analysing strain under high convergence angle conditions there are few systematic studies describing work based around the use of smaller convergence angles despite this being a common set-up. We fill in some of this gap in the literature by providing a workflow for both the experimental and analysis components of such experiments. Our case study investigates strained Gallium Arsenide nanowires with a modern direct electron detector and common microscope alignments. Three peak finding routines are compared and we provide both source code and raw data to allow others to reproduce our findings.}


\begin{document}

\section{Introduction}
Strain is an important feature of materials in a wide range of applications as it can have pronounced effects on the electrical, optical and mechanical properties of a sample. Once understood, information about the strain responses of a materials can be exploited to engineer desirable properties in a process known as `strain engineering'. This approach is widely used to improve the performance of micro-electronic devices \cite{tsutsui_strain_2019,masteghin_stress-strain_2021}, tune the opto-electronic properties of semiconductors \cite{cheng_monolayered_2018,khan_tunable_2020,peng_strain_2020} and modify the strength and hardness in metal alloy-precipitate systems \cite{gupta_strain_2015,park_strain_2018,niu_strain_2018,li_mechanical_2020}. For these methods it is essential to have accurate measurements of the strain fields present in a sample. X-Ray diffraction \cite{robinson_coherent_2009,korsunsky_mapping_1998} and micro-Raman spectroscopy \cite{li_localized_2020,ma_stressstrain_2021} are common mapping choices as they can provide highly accurate quantitative strain measurement. However the spatial resolution of these techniques is limited and in many sample the strain fields will vary at length scales on the order of nanometers. In this regime methods based on transmission electron microscopy (TEM) stand out. \\

Several TEM-based methods for strain analysis are available, each with their own strengths and weakness - for a detailed review see \cite{beche_strain_2013}. Here we will focus on approaches that collect complete diffraction patterns at many points distributed across a sample\footnote{Naming conventions differ and this technique has been described as SEND, NBED and SED in the literature but it is our belief that the similarities between these methods far exceed their differences.}. The advantage of these approaches include a wide field of view, simple set up and transparent analysis pipelines. Unfortunately, these advantages comes at the cost of lowered precision compared to other approaches. One way to improve precision is to include precession alongside scanning, to form scanning precession electron diffraction (SPED). \\

In SPED an electron beam is rastered across a sample while also undergoing a conical rocking motion (precession) to improve the sampling of reciprocal space and reduce the effects of dynamical diffraction in the final pattern \cite{vincent_double_1994,midgley_precession_2015}. By `derocking' the beam below the sample a diffraction pattern that is qualitatively similar to one that would be produced by kinematical diffraction is observed \cite{eggeman_refining_2012,eggeman_is_2010,palatinus_structure_2013}. For other application the exact nature/interpretation of the intensities under `quasi-kinematical' conditions may be of interest, but for strain mapping applications it's the increased number of Bragg peaks as well as the improved uniformity of diffraction disks that proves most beneficial \cite{pekin_optimizing_2017, yuan_lattice_2019}. \\

As scanning electron diffraction experiments create rich 4D datasets a large number of analytical approaches are possible with strain mapping being one of the most regularly applied. Recently the 4D STEM approaches have benefited from vastly improved detector quality with the advent and adoption of direct electron detectors allowing for the design of routines that require subpixel spot location \cite{nord_fast_2020,maclaren_detectorsongoing_2020,levin_direct_2021}. This development has happened in parallel with an explosion of academic open source software designed to handle and process such datasets \cite{johnstone_pyxem_2022,clausen_libertem_2020,savitzky_py4dstem_2021}. These projects allow for the transparent ingestion and analysis of TEM data and have brought new ideas to characterization workflows. While many researchers work with a converged beam (an approach sometimes called scanning CBED) our interest here is in the `quasi parallel beam' setting in which disks are well separated (see \cite{plana-ruiz_quasi-parallel_2018} for discussion of `quasi parallel beams') as this offers a simplified experimental protocol and analysis regime, while maintaining the flexibility to conduct separate investigations on the same data (eg. orientation mapping). Although different approaches have been demonstrated for analysing S(P)ED maps comparative work is limited and primarily qualitative in nature \cite{maclaren_comparing_2021}. A quantitative comparison requires analysis is done on the same strain field with multiple approaches. \\

This manuscript thus sets out to investigate the importance of the inputs provided when producing a strain map. By performing the entire mapping pipeline on a GaAs nanowire with a strained insert we seek to elucidate which elements of the process do (and don`t) have a significant effect on the final results. We describe the conditions under which three peak position finding algorithms (here: cross correlation, center of mass and curve fitting) perform well and provide the code used to generate those finding \cite{crout_workflows_2023}.\\

\section{Materials and Methods}
\subsection{Sample Preparation and Microscopy}
\begin{figure}[hbt]
    \centering
    \includegraphics[width=1\textwidth]{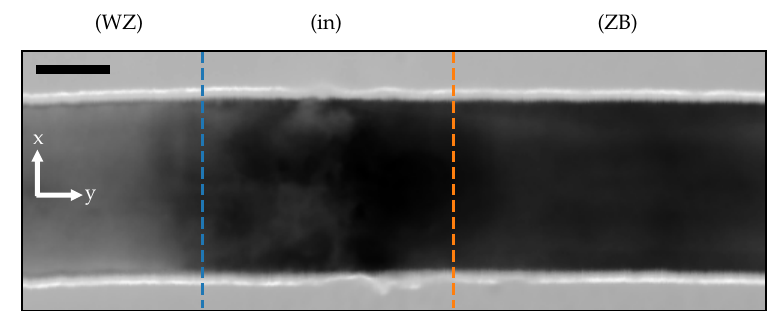}
    \caption{Virtual bright field image (VBF) of the nanowire part around the GaAsSb insert (marked by dashed lines) between zinc-blende (ZB) and wurtzite (WZ) GaAs. The nanowire is oriented to the 112/1010 zone for WZ/ZB to have similar 2-fold patterns along the scanned area. Scale bar: 50 nm.}
    \label{Fig:1}
\end{figure}

In this work we use nanowires composed of Zinc Blende (ZB) and Wurtzite (WZ) GaAs with GaAs/GaAsSb axial heterostructure. The nanowires were grown by self-catalyzation on a Si (111) substrate with the vapor-liquid-solid technique using molecular beam epitaxy. The growth and structure of these nanowires are detailed elsewhere \cite{munshi_crystal_2013}. The sample was scratched under isopropanol with a sharp diamond scraper and the droplet with detached nanowires wetted a TEM grid with an electron transparent C-film. We display the sample such that the top/left of the nanowire is the Wurtzite phase (not studied) followed by the insert and then the Zinc Blende GaAs. A contrast image demonstrating this is presented in Figure \ref{Fig:1}. The nanowire under investigation is about 4 $\mu$m long with a width of approximately 150 nm. The GaAsSb insert is of the length $\sim$ 130 nm. The growth direction was [111]/[0001] for the ZB/WZ respectively. \\

A JEOL JEM-2100F operated at an acceleration voltage of 200 kV for both the conventional TEM characterization of the specimen (which we used for verification and do not show here) and the subsequent SPED experiments. 
For the SPED work, the probe was used in nano beam diffraction (NBD) mode with a condenser aperture of 10 mm size and a spot size of 1 nm. Beam scanning was produced by a NanoMEGAS scanner and the precession alignment was conducted using the DigiSTAR software provided by the NanoMEGAS. A precession angle of 0.58$^{\circ}$, an exposure time of 10 ms and a nominal step size of 1 nm was used in all the scans. The diffraction data was acquired using a Quantum Detectors Merlin direct electron detector with a sensor size of 256x256 pixels. \\

Although multiple scans were collected and investigated, all results are presented from the scan described as `2' in Table 1 as this scan provided the best results. Strain measurements and other S(P)ED data handling were done using Python code with the open-source library Pyxem \cite{johnstone_pyxem_2022}. The full processing pipeline can be found in the notebooks available at reference \cite{crout_workflows_2023}. \\

\begin{table}[H] 
\caption{Experimental parameters for the different SPED scans collected. All scans included precession, parameters not listed here are as in the text. Investigations were conducted on all three scans, but results are only presented from scan number 2.}
\newcolumntype{C}{>{\centering\arraybackslash}X}
\begin{tabularx}{\textwidth}{CCCCCCC}
\toprule
\textbf{Scan Number} & \textbf{Region on Zone}	& \textbf{Pixel Depth} \\
\midrule
1 & Insert & 12 bit \\
2 & ZB adjacent to the insert & 12 bit \\
3 & ZB adjacent to the insert & 6 bit \\
\bottomrule
\end{tabularx}
\end{table}
\subsection{Image Processing}\label{meth:iproc}
A suitable image processing pipeline was found via a trial-and-error approach. To begin with we deal with full patterns, blocking the direct beam before cropping the relevant spots out. Each spot is then processed by passing it through a minimum filter, and retaining the largest connected region. This helps deal with the direct beam bloom in the corners of the spot images. No further image processing is applied.

\subsection{Center Finding Algorithms}

Under idealized parallel illumination conditions, a Bragg peak is infinitely sharp on a detector and provides perfect information about an associated inter-planar spacing. However, in a scanning set up, the electron beam must be converged, causing peaks to spread into disks. Each point on these disks is then subject to the noise properties of the recording system. Finding the center of these resulting disks is a crucial step in calculating strain and as such several routines have been designed and described for so called `disk finding algorithms' \cite{yuan_lattice_2019,pekin_optimizing_2017}. \\

In this work we compare three such methods. The `center of mass' approach is described first and was chosen for its intuitiveness. Following this we move onto cross correlation methods, which are commonly used and extensively described in the S(P)ED strain mapping literature \cite{pekin_optimizing_2017}. As such they provide an excellent baseline, as well as offering the interested reader a sensible starting point for considering variations with other parameters (detector setup, optical setup, post processing routine). Finally, we present a curve fitting method for finding spot centers \cite{baumann_high_2014}.

\subsubsection{Center of mass}

The centre of mass (CoM) of an object can intuitively be understood as the point from which mass `appears to act'. By analogy we can treat the intensity of a diffracting signal as its `mass' and apply a similar calculation approach. In general the center of mass/intensity for an object made of point particle is given by:
\begin{equation}\label{com}
\textbf{R}_{\text{CoM}} = \frac{ \sum_{i} \textbf{r}_{i} m_{i} } { \sum_{i} m_{i}}
\end{equation}
where $\textbf{R}_{\text{CoM}}$ is the location of CoM, and $\textbf{r}_{i}$ and $m_{i}$ are the location and weight of the constituent elements (point particles/pixels). \\

To implement this method with an individual diffraction pattern we proceed as follows: Firstly a square region containing the peak is selected. The intensity at each pixel within this square is then known and equation \ref{com} can be applied. Using this method the position of the CoM for the same disc is calculated for all the diffraction patterns in the SPED dataset.
\subsubsection{Cross correlation}
To perform cross correlation  a small reference image (usually a simulated diffraction disk of known center) is `cross correlated' with a spot from a sample. This cross correlation involves computing the value of
\begin{equation}
F(x,y) = g_{1}(x,y) \ast g_{2}
\end{equation}
where $g_{1}$ and $g_{2}$ are reference and experimental images and $\ast$ is the convolution operator for potential displacements $(x,y)$ and selecting the maximum value of $F(x,y)$. This is most easily done via a Fourier transform although we leave the internals to established Python libraries designed for this purpose \cite{pekin_optimizing_2017,walt_scikit-image_2014}. To obtain subpixel precision both the reference and the sample spot can be up-sampled. 
\subsubsection{Curve Fitting}
\begin{figure}[hbt]
    \centering
    \includegraphics[width=1\textwidth]{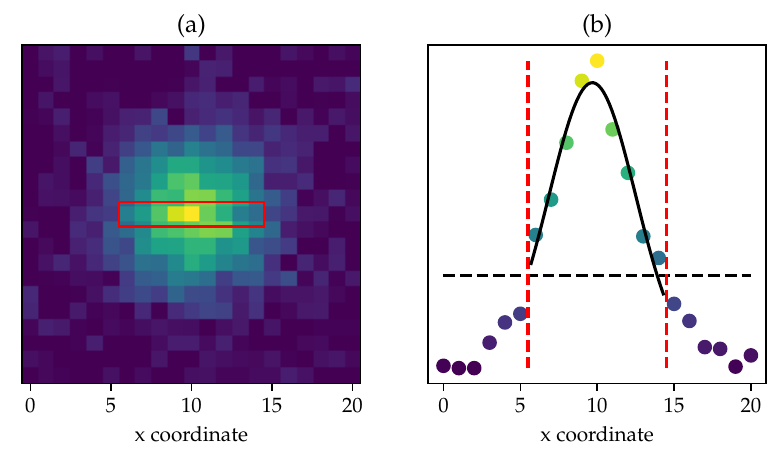}
    \caption{Illustrating the Gaussian fitting method. (a) a line trace being extracted to include only pixels with good signal to noise ratios. This is demonstrated by the red box, which is taken over a a spot with fairly substantial noise. (b) The intensity values of this trace (normalised, arbitrary units). In implementation we use a cut off ratio (dashed black line) to decide which pixels to consider. Having done this we fit a Gaussian to the included points. In this example the center is computed to be pixel 9.68, the true value is 10.0}
    \label{Fig:2}
\end{figure}

Spots from `quasi parallel' diffraction patterns show fairly consistent intensity drop off from spot center to edge. This has motivated other workers to fit models that incorporate this information. One common approach is to fit Gaussian's directly to the spot centers. The model fitting is simply least squares where the parameters are the solution to:
\begin{equation}
    \min \sum_{i} | r_{i} - M(i) |^{2}
\end{equation}
in which $i$ are the independent observations characterised by some known variable and $M$ is a model with unknown parameters. In our work we will model the intensity as following
\begin{equation}\label{eq:4}
    I(\textbf{r}) = I_{0} e^{\frac{|\textbf{r}-\mu|}{\sigma^{2}}}.
\end{equation}
for $I(\textbf{r})$ an intensity as a function of position, $I_{0}$ a fitting constant, $\mu$ an optimised center and $\sigma$ a fit width. In implementation it turns out to be far more computationally efficient to independently fit two directions and then combine the results. This is because 2D approaches require the recalculation of the values of the displacement at each step. Fitting two independent models (in which the distances can be found from a lookup table) is far quicker. Bearing this in mind, our final algorithm is:
\begin{enumerate}
    \item Cut out a square in which the spot is located (as in previous methods).
    \item Locate the highest intensity pixel in said square.
    \item Fit a 1D Gaussian to the physically relevant region of the row containing the highest intensity pixel.
    \item Fit a 1D Gaussian to the column that contains the center of our previous fit.
    \item Return the results as an $(c_{x},c_{y})$ pair.
\end{enumerate}
with implementation code available at \cite{crout_workflows_2023} and an example in Figure \ref{Fig:2}. An alternative functional fit, that of a top-hat function was also tried as these may better capture the intensity in our disks. The process is exactly as above except the function for $I(\textbf{r})$ becomes:
\begin{equation}
    I(\textbf{r}) = I_{0} \times T(\frac{|\textbf{r}-\mu|}{w}).
\end{equation}
where $T(x)$ is 1 for $x < 1$ and 0 otherwise, $w$ is a fit width parameters and then other notation is as in (\ref{eq:4}).
\subsection{Strain Calculations}\label{strain-background}
Strain ($\epsilon$) is a dimensionless tensor that measures the fractional displacement of atoms in a crystal lattice. In this, as with most TEM work, we will ignore the direction parallel to the beam, leaving us with a two dimensional problem. However even with this simplifications a number of options are available to us for converting our high precision locations of Bragg spots into a strain tensor. \\

This work adopts an algorithm that has been a staple workflow for mapping strain using SPED data for some time \cite{baumann_high_2014}. To begin with an unstrained region of the sample is found, and appropriate Bragg spots selected (these spots must be present in all patterns under consideration). The centers of these spots are then refined to subpixel precision at all relevant probe positions. The outputs are then related by a series of equations:
\begin{equation} \label{reference_deformed_strain_relationship}
    \bf{L^{i} = T L^{i}_{0}}
\end{equation}
\begin{equation}
\bf{T=RU=}
    \begin{pmatrix}
    \text{cos} \theta & \text{sin} \theta \\
    -\text{sin} \theta & \text{cos} \theta
    \end{pmatrix}
    \begin{pmatrix}
    U_{11} & U_{12} \\
    U_{12} & U_{22}
    \end{pmatrix}
\end{equation}
where $\bf{L^{i}}$ are measured spot positions in the sample, $\bf{L^{i}_{0}}$ are measured spot positions in the undeformed region, $\bf{T}$ is the mapping between the two (which varies from pixel to pixel), $\theta$ is the right handed lattice rotation about an axis pointing into the diffraction pattern and $\textbf{U}$ is the deformation tensor. From this pair of equations we can directly extract $\theta$. The strain fields ($\epsilon$) are then found by evaluating
\begin{equation}
\epsilon^{*} = \begin{pmatrix}
    \epsilon^{*}_{11} & \epsilon^{*}_{12} \\
    \epsilon^{*}_{12} & \epsilon^{*}_{22}
    \end{pmatrix} =  \bf{U} - \bf{I}
\end{equation}
where $^{*}$ indicates a value in reciprocal space. The relationship between strains in reciprocal space and real space is then given by:
\begin{equation}
    \epsilon = \frac{1}{(1+\epsilon^{*})} -1 \approx - \epsilon^{*} =  \bf{I} - \bf{U}
\end{equation}
where we have included the small strain approximation ($\epsilon^{*} \ll 1$) for comparison with other sources \cite{sankaran_multiscale_2018}. \\

\subsection{Strain extraction from peak positions}\label{sec:st}
There are now a number of ways to take our peak positions and convert them into strain values even given the constraints of the equations in section \ref{strain-background}. This is because while $\textbf{T}$ is a $2\times2$ matrix $\bf{L^{i}}$ need not be meaning that equation \ref{reference_deformed_strain_relationship} is over-determined. This `excess' of data allows for some flexibility of approach (and generally precludes the existence of a unique solution). We note that Mahr et al. were able to exploit this situation to produce a simultaneous fitting approach with success \cite{mahr_influence_2019}, but for clarity we choose here to separated our peak finding and strain mapping stages. \\

One simple approach to the problem is to sum the measured Bragg vectors into two orthogonal proxy vectors (potentially resolving vectors into components before summing). We would expect the error of such approach to be given by the standard formula for the summation of independent measurements with errors:
\begin{equation}
e_{T} = \sqrt{\sum_{i} e_{i}^{2}}
\end{equation}
for $e_{T}$ the absolute error in the (fictional) spot position and $e_{i}$ the real errors in our measurements. However, because strain is a fractional relation, the error in the strain instead is:
\begin{equation}\label{full-error}
\Delta \epsilon = \frac{\sqrt{\sum_{i} e_{i}^{2}}}{L}
\end{equation}
where $L$ is the unstrained length of the spots, we are treating the measurements in the unstrained region as errorless and the strains as small. If each spot measurement has the same error distribution this reduces to:
\begin{equation}
    \Delta \epsilon = \frac{\sqrt{N}e_{i}}{L}.
\end{equation}
As $L$ is not fixed but in fact a function of the spots chosen we can rewrite this as
\begin{equation}\label{assumption-error}
 \Delta \epsilon = \frac{\sqrt{N(c)}e_{i}}{L(c)}
\end{equation}
where $(c)$ denote a value is a function of the spots chosen. For most use cases (including this study) equation \ref{assumption-error} should suffice for estimating the error in a strain map, however if spots have markedly different error characteristics (eg. high angle spots have poorer signal-to-noise) one may wish to resort to the more involved approach of equation \ref{full-error}. 


\section{Results and Discussion}

\begin{figure}[hbt]
    \centering
    \includegraphics[width=0.95\textwidth]{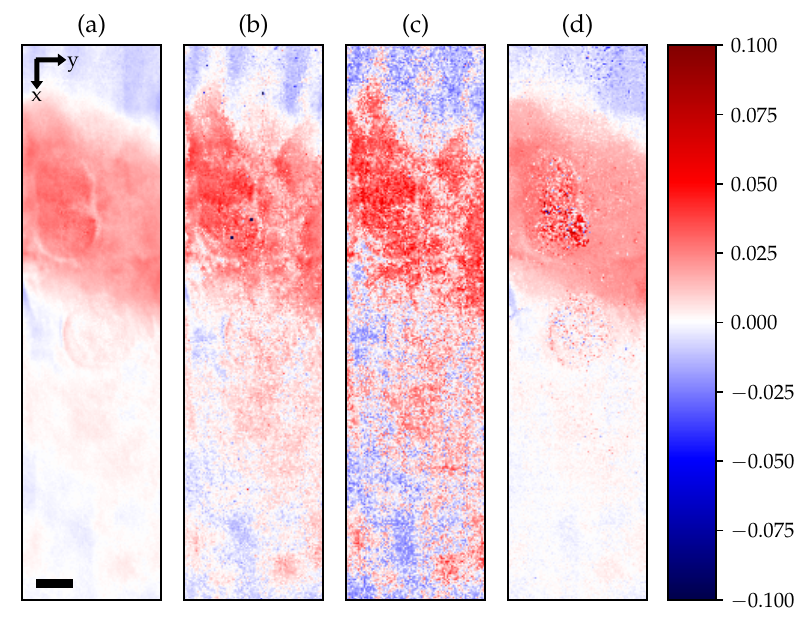}
    \caption{Strain along the nanowire ($\epsilon_{yy}$) based on the different peak shift analyzing approaches. (a) COM, (b) Gaussian fitting, (c) tophat fitting and (d) cross correlation method. Note this is a sub-region from the contrast image in Figure \ref{Fig:1}. Scalebar is 25 nm.}
    \label{Fig:3}
\end{figure}

After a suitable image processing routine was found and applied all three methods under consideration produced maps that were in qualitative agreement with one another (Figure \ref{Fig:3}). Functional fitting (using either a Gaussian or a top-hat) had significantly worse noise properties but still provided a clear indication of lattice expansion at the insert. Within the insert we see strains on the order of 2\%. The apparent compression above the insert is fictitious and due to our choice to align the microscope to the insert and be ignored as the insert causes a kink in the nanowire and the area rotates off zone. The other distinctive feature of these images are the blob shaped features which are caused by carbon contamination. \\

\begin{figure}[phbt]
    \centering
    \includegraphics[width=0.95\textwidth]{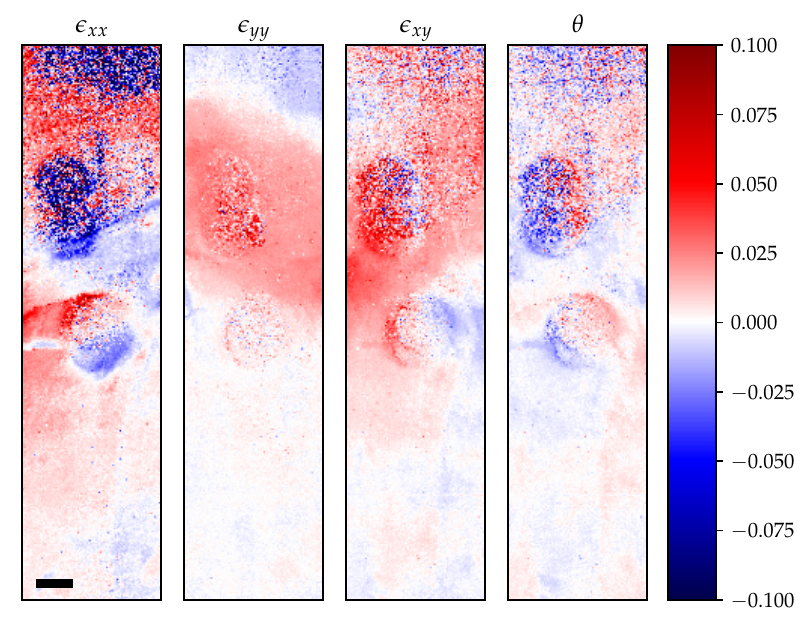}
    \caption{Showing a full set of strain maps. The poor quality of the $\epsilon_{xx}$ map demonstrates the importance of consistent, clean spots in deriving correct strain values. The method here is cross correlation and the scalebar is 25 nm.}
    \label{Fig:5}
\end{figure}

\begin{figure}[phbt]
    \centering
    \includegraphics[width=0.95\textwidth]{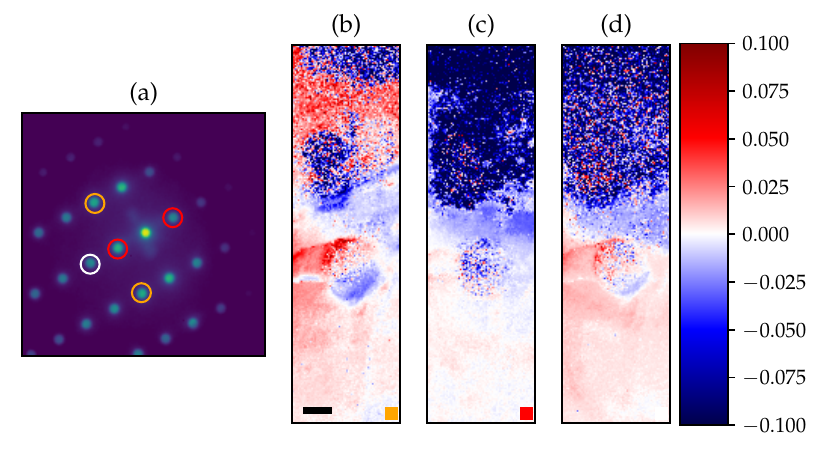}
    \caption{Demonstrating the importance of spot selection. In (a) we present a VSAED pattern, with the spots used to produce the maps (b-d) circled. Map (b) is produced by summing the two orange spots in (a) and provides an interpretable, if not particularly pleasing map of the sample. The alternatives (either subtraction of the red spots to form map (c) or the white spot alone to form (d)) produce functionally useless maps that no amount of tweaking can fix.}
    \label{Fig:7}
\end{figure}

During this investigation we found that spot quality made a significant difference to output strain maps. In Figure \ref{Fig:3} we presented $\epsilon_{yy}$ maps, as these spots are consistent throughout the scan region. In Figure \ref{Fig:5} we illustrate why this was necessary as the $\epsilon_{xx}$ component offers a significantly less clear story. The $\epsilon_{xx}$ map in Figure \ref{Fig:5} required several re-selections of spots to generate, which are in Figure \ref{Fig:7}. So while this post-collection approach did eventually succeed it would have been significantly simpler to work with our sample had our alignments allowed us to capture four orthogonal spots within the entire scan
area. \\

\begin{figure}[hbt]
    \centering
    \includegraphics[width=0.75\textwidth]{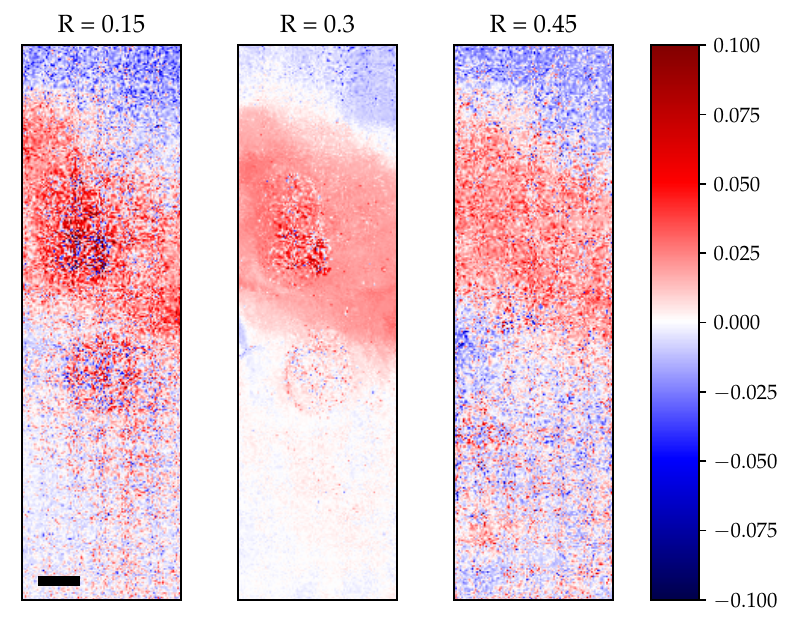}
    \caption{Demonstrating how map quality varies with image processing choice. In our set up a minimum value threshold was used (defined as $R$ the ratio of a value to the brightest pixel in a sub-pattern). We see that the maps in which $R$ allows too much ($R=0.15$) or too little ($R=0.45$) data are significantly degraded compared with our chosen parameter. These $\epsilon_{yy}$ maps are from the cross-correlation method with a scalebar of 25 nm.}
    \label{Fig:6}
\end{figure}

In what follows we focus on the more reliable $\epsilon_{yy}$ direction (due to the quality of spots in this direction) to facilitate our discussion image processing choices. As described in section \ref{meth:iproc} our approach contained four main steps (block the direct beam, cut the spots out, apply a minimum filter and then retain connected regions). While the first two steps did prove essential, they also show very limited sensitivity to the parameters selected and we will not consider them further. Keeping only connected pixels also helped to reduce erroneous results but as it contains no free parameters we suggest researcher simply toggle it on/off as they see fit. This leaves the intensity filter value as the key user defined parameter influencing the final strain map quality. In Figure \ref{Fig:6} we provide derived maps for three values of the filtering ratio and see that a fairly small change ($\pm 15\%$) induces significant degradation in the quality of the final image, which in this case appears primarily as salt and pepper noise. Beyond the map quality shown in in Figure \ref{Fig:3} there are several other considerations when selecting a center finding method.\\

The center of mass method tended to slightly underestimate strain, especially when the direct beam was included. It also produced unphysically smooth maps with noisy data which risks lulling the inattentive researcher into a false sense of security. By inspection of the lower part of Figure \ref{Fig:3} we also see that the noise properties in unstrained regions is worse than that offered by cross correlations. These deficiencies aside, the center of mass is by far the easiest to implement as well as being the cheapest (in computational time) option, allowing for rapid iterations (eg. when tuning image processing parameters).\\

In contrast, both curve fitting approaches produced deeply unpleasing maps, with significant noise. Furthermore, convergence of the underlying fitting function isn't guaranteed meaning further work is needed design and implement approaches to capturing errors. These `non-convergence timeouts' also cause the code to run far slower than the center of mass approach despite being only slightly more involved. This is disappointing as in theory a curve fitting approach most accurately captures the physics of a detector. \\

The well-established cross correlation method provided excellent results but took the most effort to coax correct results. In Figure \ref{Fig:6} we demonstrate how a small change in the image processing routine can have a dramatic impact on the quality of result obtained. While the actual algorithm takes very few parameters, all but the most aggressively cleaned spots would return complete nonsense. Due to the more involved nature of the algorithm it also proves difficult to assess a spot for defects that might have been the cause of errors. This is in stark contrast to the other methods, for which even the most cursory inspection would tend to indicate what was at fault. \\ 

Work in this field often focuses on algorithmic or technical developments at the expense of comprehensively detailing the collection and processing regimes used. Our experience with this dataset suggest that these image processing routines - as well as the initial microscope configuration and alignment - are in fact far more important. To improve the quality of our output maps we made use of several processes that are generally not well reported in the literature on strain mapping. Our choice to work only with cropped spots led to a significant decrease in required computing resources allowing for more extensive manual checks to be conducted at almost no risk of reduced accuracy. Furthermore wherever possible we used pairs of spots located at $\pm \textbf{g}$ to form our measurements vectors. This prevents artefacts associated with direct beam shift \cite{crout_methods_2023}. We also sought to use as many spots as possible to improve the SNR in the final map (see section \ref{sec:st}). \\

\section{Conclusions}
We have provided a detailed introduction to strain mapping with scanning precession electron diffraction by applying several methods to map strain in an axial heterostructured GaAs/GaASb/GaAs nanowire. This has allowed for a comparison of three popular spot finding approaches. Although it is our opinion that other components of the investigation (specifically the trio of data collection routine, image processing approach and spot selection) are greater determinant of outcome than the specific spot finding method, we can still provide some conclusions about each peak finding approach. \\

Curve fitting approaches (with either Gaussian or top-hat fits) produced comparatively poor maps. This result was surprising, as such approaches might be expected to most accurately capture the physics involved in a SPED experiment. While the center of mass method did tend to slightly underestimate strain it was by far the easiest method to implement as well producing results extremely quickly. In contrast  the well-established cross correlation method provided excellent results but was slow and fiddly to work with. All three methods required some fine user interventions and we have provided complete implementations for users to tune as they see fit \cite{crout_workflows_2023}.\\

We believe that our key finding is that a greater emphasis on the data collection component of an investigation will be productive for the vast majority of researchers. We would suggest using large precession angles with samples put as close to zone-axis as possible. Alongside this having two pairs of orthogonal spots produces significantly better results. One potential difficulty is sample tilt, but if this is not too severe through the region of interest we believe any correct implementation of the spot finding methods detailed in this paper will produce good results.
\newpage
\vspace{6pt} 



\authorcontributions{Conceptualization, I.N.P, D.C, T.I.T and A.T.J.H; methodology, all listed authors; software and formal analysis P.C; data curation, D.C, I.N.P and T.I.T; writing---original draft preparation, P.C and D.C; writing---review and editing, P.C and A.T.J.H; visualization, P.C and I.N.P.; supervision, project administration and funding acquisition: A.T.J.H and P.A.M. All authors have read and agreed to the published version of the manuscript.}

\funding{This project formed a component of P.C's doctoral research which was funded by the EPSRC and NPL. All authors acknowledge the Research Council of Norway for their support through the Norwegian Center for Transmission Electron Microscopy: NORTEM (Grant no. 197405).}

\dataavailability{The raw data associated with this manuscript can be found at \cite{crout_data_2023}. The Python scripts used to create many of the results are freely available at \cite{crout_workflows_2023}.} 







\appendixtitles{no} 

\begin{adjustwidth}{-\extralength}{0cm}

\reftitle{References}


\bibliography{crout_et_al.bib}

\end{adjustwidth}
\end{document}